\documentclass[epj]{svjour}
\usepackage{graphics}
\begin{document}
\title{Chiral Symmetry and Hyperfine Meson Splittings}
\author{Felipe J. Llanes-Estrada
\inst{1}
\thanks{\emph{On leave at University
of Tuebingen, Inst. fuer Theoretische Physik, auf der Morgenstelle 14,
 D-72076  Tuebingen, Germany.} }
\and Stephen R. Cotanch \inst{2}
\and Adam P. Szczepaniak \inst{3}
\and Eric S. Swanson \inst{4}
% etc
}                     % Do not remove
\institute{Depto. de F\'{\i}sica Te\'orica I,  Univ. Complutense, 28040
Madrid, Spain.
\and Department of Physics, North Carolina State
University, Raleigh NC 27695-8202 USA.
\and Department of Physics and Nuclear Theory Center, Indiana
University Bloomington, IN 47405-4202 USA.
\and
Department of Physics and Astronomy, University of
Pittsburgh, Pittsburgh, PA 15260
 and Jefferson Lab, 12000 Jefferson Ave, Newport News, VA 23606 USA.
}
\date{Oct. 31, 2003}
\abstract{
We briefly review  theoretical calculations for the
pseudoscalar-vector meson hyperfine splitting with no open
flavor and also report a many body field theoretical effort to assess the
impact of  chiral symmetry in the choice of effective potentials for
relativistic quark models. Our calculations predict the missing $\eta_b$
meson to have mass near 9400 $MeV$. The radial excitation $\eta_c(2S)$ is
in agreement with the measurements of the Belle and most recently Babar
collaborations.
 % end of PACS codes
} %end of abstract
\maketitle
%%%%%%%%%%%%%%%%%%%%%%%%%%%%%%%%%%%%%%%%%%%%%%%%%%%%%%%%%%%%%%%%%%

Shortly after the discovery of the ${\rm J}/\psi$ it was understood
that a rich spectroscopy of new mesons awaited classification.
In this task the constituent quark model was a useful tool  providing
a simple periodic table where spectra and various radiative decays
could be correlated with the help of a modest number of parameters.
In this picture vector mesons are a $q \overline{q}$ pair, in
an S or D wave, with spins parallel giving total angular momentum
$J=1$.
Pseudoscalar mesons correspond to the $J=0$ ground state with
S-wave $q \overline{q}$
pairs  spins antialigned.
Ignoring the D-wave component, the only difference
between both systems is the relative spin alignment.
Any spectroscopic mass splitting can conveniently be
incorporated in the quark model with a term,
$A \vec{\sigma_1}\cdot \vec{\sigma_2}$
that is reminiscent of the electron-nucleus spin-spin coupling,
hence the name ``hyperfine''.
This was immediately noted  by  Appelquist {\it et al.}~\cite{deRujula}
who predicted a  charmonium splitting,
$\Delta M(J/\psi-\eta_c)$, of about $65 \ MeV$. They extracted the
amplitude $A$ by estimating the $J/\psi$ electron-positron width,
$\Gamma_{e^-e^+}$, to be $4 \ keV$. Using the currently
accepted value of $5.3\ keV$, the splitting would be about $84 \ MeV$,
or about a factor of 2 smaller than the accepted experimental value of
$3097-2980 \simeq 120\ MeV$.
The need for a confining potential \cite{cornell} was soon understood
and calculations (by Appelquist and Politzer, and independently
Schnitzer \cite{schnitzer}) including a confining strength yielded a
larger splitting ($40-80 \ MeV$) than purely Coulombic potentials
($15-20\ MeV$).

In retrospective we see that many of the early models utilized
scalar confining potentials, which provided a good spin-orbit coupling
and radial excitations, but underestimated the  hyperfine splittings.
In the early eighties, and with the $\eta_c$ experimental state now known,
this splitting became a benchmark for new model
calculations \cite{isgur,McClary,Eichten,igi} which now also predicted the
corresponding splitting in bottomonium. The variation in
these predictions is summarized in Table \ref{tab:1}.
Subsequently, further progress was achieved through improved,
renormalized non-relativistic perturbative QCD calculations (NRPQCD)
\cite{yndurain,brambilla} which
described bottomonium as a  non-relativistic system. However,
the calculated radii of most $b\overline{b}$ states are too large
indicating that a Coulombic description, where the relativistic
splittings scale linearly with the quark mass, is not reliable and that strong
interactions still induce important corrections at this scale
\cite{Leutwyler}. Nevertheless, approximate  ground
state descriptions are feasible and useful for extracting $c$ and $b$
quark masses.

Non-perturbative lattice calculations with large error bars have also been
performed
\cite{Davies1,Davies2}  for bottomonium which yield about half, or less,
the hyperfine  splitting exhibited in charmonium.  This again
indicates the system is not fully Coulombic since the splitting
is not proportional to the quark mass.

\begin{table}
\caption{Predictions for the splitting between vector
and pseudoscalar $b\overline{b}$ mesons. CQM stands for Constituent Quark
Model. Units are $MeV$.}
\label{tab:1}
\begin{tabular}{lll}
\hline\noalign{\smallskip}
Date, Authors & Model & Splitting \\
\noalign{\smallskip}\hline\noalign{\smallskip}
1983 Godfrey \& Isgur  & CQM & 60 \\
1983 McClary \& Byars  & CQM & 101  \\
1985 Igi \& Ono & CQM Coulombic     &  60  \\
1985 Igi \& Ono & CQM log running  &  90  \\
1989 Song & CQM                     &  55  \\
1994 Eichten \& Quigg & CQM Cornell & 141 \\
1994 Eichten \& Quigg & CQM various & 87/65/64 \\
1994(98) Davies {\it et al.}& Lattice & 30-50 \\
1998 Pineda \& Yndurain & NRQCD      & 47(20) \\
2000 Lengyel {\it et al.}& CQM        & 46 \\
2003 Ebert  {\it et al.} & CQM    & 60 \\
\noalign{\smallskip}\hline
\end{tabular}
\end{table}

Extending this analysis accurately to the $\pi$-$\rho$ system is not
currently feasible for either the perturbative or lattice approaches. Thus
one still relies on constituent models where the hyperfine splitting has
a $1/M^2$ dependence on the  constituent quark mass \cite{isgur}. This
can describe the large  $\pi$-$\rho$ splitting but
not simultaneously the hadron scattering phase shifts \cite{bicudo1}.
On the other hand, we know that the pion's mass is very low because
of its  Goldstone boson nature from spontaneous chiral symmetry
breaking. Thus it is natural to seek a field-theoretical formulation of
the quark model which  implements chiral symmetry  consistently.
Such an approach would predominantly attribute the hyperfine splitting
in light mesons to chiral symmetry. This permits using a more moderate
hyperfine potential to then decribe
the smaller splittings which are exhibited in light meson excited
states and
heavy mesons, both of which are not governed by chiral symmetry. Thus we
consider the Hamiltonian (inspired in Coulomb gauge QCD)
\begin{eqnarray} \label{hamiltonian}
H_{eff} &=& T+V_{C}+V_{T}   \\
T &=& \int d{\bf x} \Psi^\dagger ({\bf x}) ( -i \mbox{\boldmath
$\alpha$\unboldmath}
\cdot
\mbox{\boldmath   $\nabla $\unboldmath} + m_q \beta ) \Psi ({\bf x})   \\
V_C &=& -\frac{1}{2} \int d{\bf x} d{\bf y}
\rho^a ({\bf x})
\hat{V}(\arrowvert {\bf x}-{\bf y} \arrowvert )
\rho^a ({\bf y})   \\
V_T &=& \frac{1}{2} \int d{\bf x} d{\bf y}
J^a_{i} ({\bf x})
J^a_{j} ({\bf y}) \times  \nonumber \\
&&
\left( \delta_{ij} - \frac{\nabla_i \nabla_j}{\nabla^2} \right)_{\bf x}
 \hat{U}(|{\bf x}-{\bf y}|) \ .
\end{eqnarray}
Here
$\rho^a =  \Psi^\dagger T^a \Psi$ and ${\bf J} ^a = \Psi^\dagger
\mbox{\boldmath$\alpha$\unboldmath} T^a \Psi$ are the quark color density and
current, respectively. This Hamiltonian has been diagonalized
previously~\cite{early} for
$V_T=0$ in the Bardeen-Cooper-Schrieffer (BCS) approximation for the
vacuum. These
earlier studies of the gap equation determined that the dynamical chiral
symmetry
breaking from only a longitudinal potential is relatively small and yields
a  low condensate $\langle \overline{\Psi} \Psi \rangle_0 \simeq -(100
\  MeV)^3$.
On the opposite limit,  calculations for high quark masses using the
Tamm-Dancoff (TDA) and Random Phase (RPA) approximations for both
harmonic oscillator  \cite{orsay2,bicudo2} and  linear potentials
\cite{LC} produce almost degenerate pseudoscalar and vector meson
ground states.
 They are thus unable to describe the charmonium
hyperfine  splitting  although the RPA can reproduce the
$\pi$-$\rho$  splitting by sufficiently lowering the quark mass according to
Thouless  theorem.
More recently  a study~\cite{ases} implementing chiral symmetry used
$V_C=0$ and
a contact potential for
$V_T$  to obtain a link with transverse one-gluon exchange, which is suppressed
in our approach by the large gluon mass gap \cite{ssjc}. Because that model
does not include radial excitations or confinement,
we have generalized~\cite{inprep} the treatment  by employing both a
Coulomb  instantaneous interaction and a transverse hyperfine potential.
For the longitudinal Coulomb interaction we utilize a potential
derived~\cite{sspot} from QCD through  a BCS truncation of the gluon sector,
represented in momentum  space by
\begin{eqnarray} \label{SSpotential} \nonumber
\hat{V}(p)&=& C(p) \equiv -\frac{8.07}{p^2} \, \frac{{\rm log}^{-0.62} \left(
\frac{p^2}{m_g^2}  +0.82 \right) }{{\rm log}^{0.8}\left(
\frac{p^2}{m_g^2} +1.41\right)  } \ \ {\rm for} \ p >m_g \\
\hat{V}(p) &=& -\frac{12.25 \, m_g^{1.93}}{p^{3.93}} \ \ {\rm for} \
p <m_g \
.
\end{eqnarray}
This is numerically close to the standard Coulomb + linear potential.
The transverse potential, due to non-explicit Lorentz covariance in
Coulomb gauge QCD, can be different. Since this term has not been studied
theoretically, we proceed phenomenologically and choose the same Coulomb
tail as in Eq. (\ref{SSpotential}).  It is then matched at low momentum to a
Yukawa  representing a massive gluon exchange which emerges
from intermediate hybrid states in the Fock space
truncation. Thus we take
\begin{eqnarray} \label{yukawa}
\hat{U}(p)&=& C(p) \ \ {\rm for} \ p>m_g \\ \nonumber
\hat{U}(p)&=& -\frac{C_h}{p^2+m_g^2}   \ \ {\rm for} \ p <m_g \ .
\end{eqnarray}
The constant $C_h$ matches the potential continuously at the $m_g$ scale.
Thus the only free potential parameter is $m_g$ which
determines simultaneously the strength of the confining term and the
logarithmic one-loop running of both $\hat{U}$ and $\hat{V}$. We adopt
$m_g=600 \ MeV$ and  investigate alternative transverse potentials in
a more detailed publication~\cite{inprep}.

Calculating the resulting gap equation at zero quark mass we find a sizeable
increase of the BCS quark condensate,  to  $-(178 \ MeV)^3$, which is
now closer to the phenomenologically accepted values (this
quantity is sensitive to the high energy behaviour of the potential as
previously noted by Lagae~\cite{Lagae}). In the chiral
limit the calculated pion mass is effectively zero (numerically a fraction
of an
$MeV$) and the $\rho$ mass is about $780 \ MeV$.
For the vector mesons we include coupled $S$ and $D$ wave channels,
since the Hamiltonian of Eq. (\ref{hamiltonian}) contains a tensor
interaction.

\begin{figure}
\resizebox{0.45\textwidth}{!}{
\includegraphics{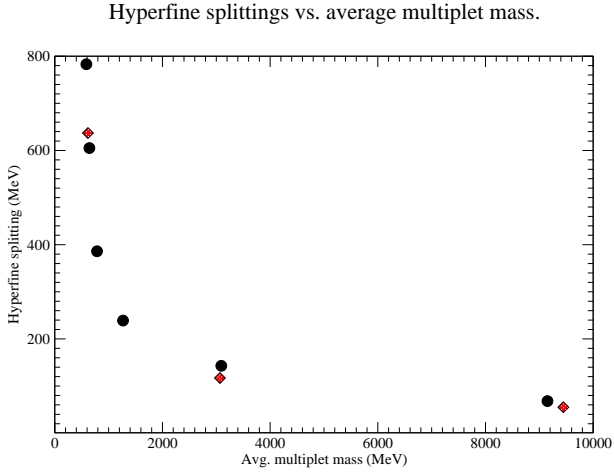}}
\vspace{0.2cm}
\caption{\label{fig1} RPA ground state hyperfine splittings,
$M(1^{--})-M(0^{-+})$, versus the average
multiplet mass $(3M(1^{--})+M(0^{-+}))/4$
(circles). The three diamonds
represent the observed $\pi$-$\rho$ and
$\eta_c$-$J/\psi$ and the
NRPQCD $\eta_b$-$\Upsilon$ splittings.
\vspace{0.5cm}}
\end{figure}

Upon increasing the quark mass, the
pion mass growes rapidly in the RPA whereas the $\rho$ mass only
slowly increases yielding  the hyperfine
splitting plotted in Fig. \ref{fig1} for various meson masses.
This figure presents
our preliminary results and reflects the success of this approach
which incorporates
chiral symmetry and is
simultaneously applicable to a wide range of quark masses.

For the same model parameters, we also predict the mass
of the missing $\eta_b$ state, a most important issue in
hadronic spectroscopy~\cite{godfrey}.  We concur
with NRPQCD and lattice studies but predict a slightly larger splitting,
(see below) of  about $60 \ MeV$.
Subtracting this from the $\Upsilon(9460)$ mass yields $\eta_b(9400)$. This
decreasing hyperfine strength trend with increasing quark mass (see
Fig. \ref{fig1})  indicates that the potential is not yet scaleless. Note that
in both PQCD and our approach (see Eq. \ref{SSpotential}) a hadron scale
appears
logarithmically in the coupling constant.
Also for bottomonium there is a small difference between
the RPA and TDA hyperfine splittings since the TDA $\eta_b$ mass is
about $30 \ MeV$ lower than in the RPA. While insignificant when compared
to the $\Upsilon (9460 \  MeV)$ mass, it should be accurately included
when evaluating a small hyperfine splitting.
Non-chiral preserving models, such as those
based on Schr\"odinger's equation, will thus underestimate
the splitting by at least this $30 \ MeV$. Although this is
currently comparable to
the quoted errors in both NRPQCD and lattice calculations ($20-30 \ MeV$),
it may become an issue in the future.

Finally, it is noteworthy that our approach naturally extends
to  radial excitations. For the
$\psi(2S)-\eta_c(2S)$ splitting we obtain $56 \ MeV$, in agreement
with the BELLE \cite{belle} result. {\it Note added: the BABAR
\cite{babar} collaboration reports a possible detection of the
$\eta_c(2S)$ corresponding to a hyperfine splitting of $55(4)\ MeV$.
The agreement is encouraging. }

This work was supported by Spanish grants
FPA 2000-0956, BFM 2002-01003
(F.L-E.) and the U. S. Department of Energy
grants DE-FG02-97ER41048
(S.C.), DE-FG02-87ER40365 (A.S.) and
DE-FG02-00ER41135, DE-AC05-84ER40150
(E.S.).

%%%%%%%%%%%%%%%%%%%%%%%%%%%%%%%%%%%%%%%%%%%%%%%%%%%%%%%%%%%%%%%%%%%

\end{document}